\documentclass[10pt,twocolumn,brazil,aps]{revtex4}
\usepackage[english,portuguese]{babel}
\usepackage[T1]{fontenc}
\usepackage[latin9]{inputenc}
\usepackage{color}
\usepackage{amsmath}
\usepackage{amssymb}
\usepackage{graphicx}
\usepackage[unicode=true,pdfusetitle,
 bookmarks=true,bookmarksnumbered=false,bookmarksopen=false,
 breaklinks=false,pdfborder={0 0 0},backref=false,colorlinks=true]
 {hyperref}
\usepackage{url}

\makeatletter

\@ifundefined{textcolor}{}
{%
 \definecolor{BLACK}{gray}{0}
 \definecolor{WHITE}{gray}{1}
 \definecolor{RED}{rgb}{1,0,0}
 \definecolor{GREEN}{rgb}{0,1,0}
 \definecolor{BLUE}{rgb}{0,0,1}
 \definecolor{CYAN}{cmyk}{1,0,0,0}
 \definecolor{MAGENTA}{cmyk}{0,1,0,0}
 \definecolor{YELLOW}{cmyk}{0,0,1,0}
}

\hypersetup{urlcolor=blue, citecolor=blue, linkcolor=blue}

\makeatother

\begin{document}

\title{Entendendo a Entropia de von Neumann \\ \small{(Understanding von Neumann's entropy)}}

\begin{abstract}
Revisamos os postulados da mec\^anica qu\^antica necess\'arios para discutir
a entropia de von Neumann, a introduzimos como uma generaliza\c{c}\~ao da
entropia de Shannon e propomos um jogo simples que facilita o
entendimento do seu significado f\'isico. \\
\textbf{Palavras chave:} Mec\^anica qu\^antica, Operador densidade, Entropia de von Neumann \\

\vspace{0.01mm}

We review the postulates of quantum mechanics that are needed to discuss the
von Neumann's entropy. We introduce it as a generalization of Shannon's
entropy and propose a simple game that makes easier understanding its physical
meaning. 
\end{abstract}

\maketitle

\section{Introdu\c{c}\~ao}

A entropia \'e uma fun\c{c}\~ao de estado de fundamental import\^ancia em f\'isica,
em teoria da informa\c{c}\~ao e em v\'arias outras \'areas da ci\^encia esteadas
nestas duas. A entropia \'e, em geral, primeiramente introduzida
de maneira fenomenol\'{o}gica no contexto da termodin\^{a}mica \cite{Van Ness}. Em mec\^{a}nica estat\'{i}stica, a entropia \'e obtida usando teoria de probabilidades em conjun\c{c}\~ao com as leis din\^{a}micas que governam o movimento dos constituintes microsc\'{o}picos do sistema \cite{Salinas}. Investiga\c{c}\~oes sobre a entropia geraram um longa e interessante hist\'oria, deixando pelo caminho algumas controversas e quest\~oes em aberto. Para mais discuss\~oes sobre esses aspectos hist\'{o}ricos, fundamentais e tamb\'em sobre temas atuais de pesquisa relacionados ao assunto, indicamos ao leitor as seguintes refer\^encias bibliogr\'aficas \cite{Neumann,Goldstein,Jaynes1,Jaynes2,Jaynes3,Wehrl,Petz,Popescu,Popescu2,Lieb,Baumgartner,Jarzynski,Serra}.
Neste artigo focaremos nossa aten\c{c}\~ao na defini\c{c}\~ao para a entropia fornecida no contexto do formalismo matem\'{a}tico da mec\^anica qu\^{a}ntica (MQ). Na
sequ\^encia apresentamos uma breve revis\~ao sobre a MQ, com \^enfase especial
em aspectos necess\'arios e motivacionais para a introdu\c{c}\~ao da entropia
de von Neumann como uma medida de falta de conhecimento, ou de incerteza,
no contexto de prepara\c{c}\~oes e testes. Por fim, introduzimos um jogo
simples que ajuda na compreens\~ao dessa fun\c{c}\~ao.

\section{Postulados da mec\^anica qu\^antica}

\subsection{Postulados dos estados}

Para descrever um certo sistema f\'isico, come\c{c}amos associando a ele
um \textit{espa\c{c}o de Hilbert}, que, para nossas propostas nesse artigo, nada
mais \'e do que o espa\c{c}o vetorial formado por todas as listas com $n$
n\'umeros complexos, conhecido como $\mathbb{C}^{n}$, com uma fun\c{c}\~ao
produto interno bem definida \cite{Griffiths,Piza}. O pr\'oximo passo
\'e associar cada estado, ou configura\c{c}\~ao, poss\'ivel do sistema com um
\'unico vetor de $\mathbb{C}^{n}$. Aqui usaremos a nota\c{c}\~ao de Dirac
para estes vetores, ou seja, para $\vec{z}\in\mbox{\ensuremath{\mathbb{C}}}^{n}$,
denotaremos
\begin{equation}
|z\rangle\equiv\vec{z}=\begin{bmatrix}z_{1}\\
\vdots\\
z_{n}
\end{bmatrix},
\end{equation}
ou seja, $|z\rangle$ \'e uma matriz coluna. \'E frequente tamb\'em a utiliza\c{c}\~ao
da adjunta (transposta conjugada) dessa matriz, que escreveremos como
\begin{equation}
\langle z|\equiv|z\rangle^{\dagger}=\begin{bmatrix}z_{1}^{*} & \cdots & z_{n}^{*}\end{bmatrix},
\end{equation}
que \'e uma matriz linha.

A \textit{fun\c{c}\~ao produto interno} entre dois vetores quaisquer $|z\rangle,|w\rangle\in\mathbb{C}^{n}$,
comentada anteriormente, \'e denotada por $\langle z|w\rangle$ e \'e
definida como o seguinte produto de matrizes
\begin{equation}
\langle z|w\rangle\equiv|z\rangle^{\dagger}|w\rangle=\begin{bmatrix}z_{1}^{*} & \cdots & z_{n}^{*}\end{bmatrix}\begin{bmatrix}w_{1}\\
\vdots\\
w_{n}
\end{bmatrix}=\sum_{i=1}^{n}z_{i}^{*}w_{i}.
\end{equation}

Note que se trocarmos a ordem das matrizes no produto interno da \'ultima
equa\c{c}\~ao n\~ao teremos mais um n\'umero complexo, mas sim uma matriz $n\mathrm{x}n$:
\begin{equation}
|w\rangle|z\rangle^{\dagger}=|w\rangle\langle z|=\begin{bmatrix}w_{1}z_{1}^{*} & \cdots & w_{1}z_{n}^{*}\\
\vdots & \ddots & \vdots\\
w_{n}z_{1}^{*} & \cdots & w_{n}z_{n}^{*}
\end{bmatrix}.
\end{equation}
Este tipo de produto de matrizes, conhecido como \textit{produto externo},
ser\'a utilizado na sequ\^encia para discutirmos a descri\c{c}\~ao de observ\'aveis
em MQ e tamb\'em quando introduzirmos o operador (ou matriz) densidade
para descrever o estado de um sistema.

\subsection{Postulado das medidas}

Este postulado nos fornece uma regra para calcular as probabilidades
(as chances) de obtermos um certo valor de um observ\'avel quando montamos
um experimento capaz de discernir entre seus poss\'iveis valores. Do
ponto de vista matem\'atico, um observ\'avel \'e representado por um operador
linear e hermitiano $\hat{O}$, o que garante que este pode ser escrito
na forma de uma decomposi\c{c}\~ao espectral: $\hat{O}=\sum_{i=1}^{n}o_{i}|o_{i}\rangle\langle o_{i}|,$
com seus autovalores sendo n\'umeros reais ($o_{i}\in\mathbb{R}$) e
seus autovetores, normalizados, correspondentes a autovalores diferentes
sendo ortogonais ($\langle o_{i}|o_{j}\rangle=\delta_{ij}$). Por
simplicidade, consideramos somente espectros discretos e n\~ao degenerados
\cite{Arfken}. 

Na MQ n\~ao h\'a, em geral, como prever de antem\~ao qual valor $o_{i}$
ser\'a obtido em uma medida (em um teste) do observ\'avel $\hat{O}$ \cite{Bell}.
O que podemos computar \'e a probabilidade condicional de obter $o_{i}$
quando o sistema foi preparado no estado $|z\rangle$. Para esse prop\'osito
usamos a \emph{regra de Born}: 
\begin{equation} 
\mathrm{Pr}(o_{i}|\vec{z})\equiv|\langle z|o_{i}\rangle|^{2}. 
\end{equation}

Voc\^es devem estar se perguntando: Mas como determinamos que o sistema
est\'a preparado no estado $|z\rangle$? Aqui entra o papel fundamental
do conceito de \emph{prepara\c{c}\~oes e testes}, enfatizado por Asher Peres
em seu livro sobre MQ \cite{Peres}. Um padr\~ao importante que percebemos
observando o que ocorre em experimentos \'e que se fizermos medidas
subsequentes, uma imediatamente depois da outra, obteremos sempre
o mesmo valor para o observ\'avel. Ent\~ao, quando dizemos que o estado
no qual o sistema foi preparado \'e $|z\rangle$, queremos dizer que
um certo observ\'avel foi medido e que se obteve o valor dele correspondente
ao seu autovetor $|z\rangle$. Ou seja, \emph{o estado qu\^antico de
um sistema representa o conhecimento (a informa\c{c}\~ao) que possu\'imos
acerca de sua prepara\c{c}\~ao}. Usando essa informa\c{c}\~ao, e a regra de Born,
podemos fazer previs\~oes sobre medidas (testes) de outros observ\'aveis.

\subsection{Princ\'ipio de superposi\c{c}\~ao}

Este princ\'ipio \'e uma consequ\^encia imediata da estrutura do espa\c{c}o
de estados usado na MQ. Note que o conjunto de autovetores de qualquer
observ\'avel de um certo sistema f\'isico forma uma base para $\mathbb{C}^{n}$,
pois qualquer vetor desse espa\c{c}o vetorial pode ser escrito como uma
combina\c{c}\~ao linear (uma superposi\c{c}\~ao) desses vetores, i.e., 
\begin{equation}
|z\rangle=\sum_{i=1}^{n}c_{i}|o_{i}\rangle,
\end{equation}
com $c_{i}=\langle o_{i}|z\rangle\in\mathbb{C}$. 
Para um n\'umero muito
grande de medidas do observ\'avel $\hat{O}$, as probabilidades fornecidas
pela regra de Born, $|\langle z|o_{i}\rangle|^{2}$, nos informam
qual \'e a frequ\^encia relativa com que os autovalores correspondentes
de $\hat{O}$ ser\~ao obtidos. Vemos assim que sempre que existir dois
ou mais coeficientes n\~ao nulos na superposi\c{c}\~ao, ou seja, sempre que
o estado preparado n\~ao for um autovetor do observ\'avel, existir\~ao tamb\'em
duas ou mais probabilidades n\~ao nulas. Isso implica que, no formalismo
da MQ, h\'a obrigatoriamente incerteza sobre qual ser\'a o resultado obtido
em uma dada realiza\c{c}\~ao do experimento para medir o valor de $\hat{O}$.
Essa incerteza, que chamaremos de \emph{incerteza qu\^antica,} tem sua
origem na exist\^encia de observ\'aveis incompat\'iveis em MQ. Estes observ\'aveis
s\~ao representados por operadores que n\~ao comutam e que, portanto,
n\~ao compartilham uma mesma base de autovetores \cite{Griffiths}.
E \'e isso que possibilita, atrav\'es da medida de um observ\'avel $\hat{Z}$,
a prepara\c{c}\~ao de um estado $|z\rangle$ que n\~ao \'e autovetor de $\hat{O}$.

\subsection{Operador densidade}

\begin{figure}
\includegraphics[scale=0.6]{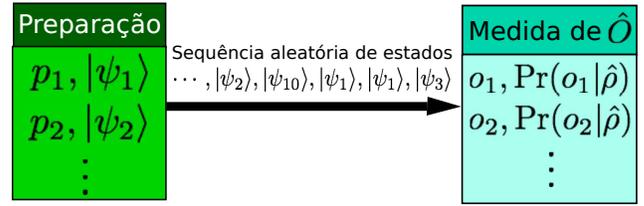}
\protect\caption{Se n\~ao conhecemos com certeza qual \'e o estado de um membro de um 
de part\'iculas, mas sabemos que a chance do seu estado ser $|\psi_{j}\rangle$
\'e $p_{j}$, podemos usar o operador densidade, $\hat{\rho}$, para
calcular a probabilidade de obter o valor $o_{i}$ em medidas de um
observ\'avel $\hat{O}$ desse sistema.}
\label{fig_ensamble}
\end{figure}

\begin{figure}
\includegraphics[scale=0.5]{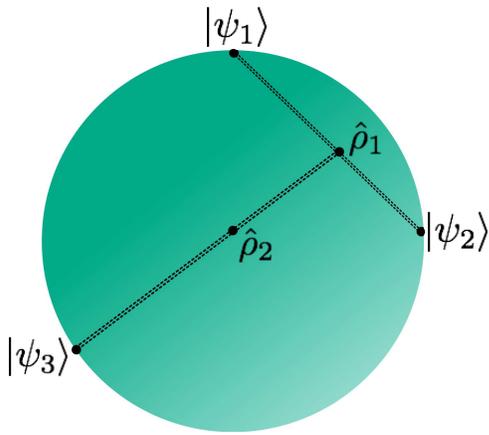}
\protect\caption{Espa\c{c}o de estados: Estados puros (vetores de estado) est\~ao na borda
de uma hiperesfera com raio igual a um. Por sua vez, os operadores
densidade (que s\~ao misturas estat\'isticas de vetores de estado) est\~ao
no interior dessa hiperesfera ($\hat{\rho}_{1}=p_{1}|\psi_{1}\rangle\langle\psi_{1}|+p_{2}|\psi_{2}\rangle\langle\psi_{2}|$
e $\hat{\rho}_{2}=\hat{\rho}_{1}+p_{3}|\psi_{3}\rangle\langle\psi_{3}|$).}
\label{fig_state_space}
\end{figure}

Consideremos a possibilidade de exist\^encia de incerteza em rela\c{c}\~ao
ao estado no qual o sistema foi preparado. Por exemplo, considere
que algu\'em, ou algo, com acesso a v\'arias c\'opias id\^enticas de um sistema
f\'isico prepare essas c\'opias em um conjunto qualquer de estados $\{|\psi_{j}\rangle\}_{j=1}^{m}$
de $\mathbb{C}^{n}$ com probabilidades respectivas $\{p_{j}\}_{j=1}^{m}$.
Claro, se nenhum sistema for perdido, $\sum_{j=1}^{m}p_{j}=1$. Estes
estados n\~ao precisam ser autovetores de observ\'aveis compat\'iveis. Nossa
tarefa \'e encontrar o objeto matem\'atico que pode ser usado para obter
a express\~ao mais simples poss\'ivel para a probabilidade de medir o
valor $o_{i}$ de um observ\'avel $\hat{O}$ desse sistema. Sabemos
que, para cada medida, o sistema foi preparado em um dos estados puros
desse ensemble (desse conjunto) e que, por conseguinte, a regra de Born pode ser aplicada
ao sub-ensemble preparado no estado $|\psi_{j}\rangle$. Ou seja,
a probabilidade de medirmos $o_{i}$ no conjunto de sistemas preparados
no estado $|\psi_{j}\rangle$ \'e: $\mathrm{Pr}(o_{i}|\vec{\psi}_{j})=|\langle\psi_{j}|o_{i}\rangle|^{2}$.
Com isso, a probabilidade que queremos calcular, que \'e a chance de
medida de $o_{i}$ considerando todo o ensemble, pode ser escrita
como a m\'edia dessas probabilidades calculada usando as probabilidades
de prepara\c{c}\~ao $p_{j}$ \cite{prob,DeGroot}:
\begin{eqnarray}
\mathrm{Pr}(o_{i}|\{p_{j},\vec{\psi}_{j}\}) & = & \sum_{j=1}^{m}p_{j}\mathrm{Pr}(o_{i}|\vec{\psi}_{j})=\sum_{j}p_{j}\langle\psi_{j}|o_{i}\rangle\langle o_{i}|\psi_{j}\rangle\nonumber \\
 & = & \langle o_{i}|\sum_{j}p_{j}|\psi_{j}\rangle\langle\psi_{j}|o_{i}\rangle=\langle o_{i}|\hat{\rho}|o_{i}\rangle,
\end{eqnarray}
onde definimos o \emph{operador densidade}, que descreve completamente
o estado do sistema, como a seguinte mistura estat\'istica dos estados
do ensemble: 
\begin{equation}
\hat{\rho}=\sum_{j=1}^{m}p_{j}|\psi_{j}\rangle\langle\psi_{j}|.
\end{equation}
Este processo de dedu\c{c}\~ao de $\hat{\rho}$ est\'a ilustrado na Fig. \ref{fig_ensamble}.
Uma representa\c{c}\~ao esquem\'atica do espa\c{c}o de estados pode ser vista
na Fig. \ref{fig_state_space}.

Note que, para um ensemble com um n\'umero $N$ muito grande de elementos,
$p_{j}N$ fornece o n\'umero de part\'iculas preparadas no estado puro
$|\psi_{j}\rangle$. Outra observa\c{c}\~ao pertinente aqui \'e que nossa
incerteza sobre a prepara\c{c}\~ao do estado do sistema n\~ao tem nada de
especial no que concerne a MQ. Isso porque a informa\c{c}\~ao sobre qual
estado foi preparado, para cada membro do ensemble, est\'a presente
na mem\'oria cl\'assica do preparador. Por isso, essa incerteza \'e chamada
de \emph{incerteza cl\'assica}.

\section{Entropia de von Neumann}

Por raz\~oes did\'{a}ticas, abdicaremos da sequ\^encia hist\'orica dos fatos e introduziremos a entropia de Shannon antes da entropia de von Neumann.
Para isso, come\c{c}aremos respondendo a seguinte pergunta: Se quisermos
associar uma medida de informa\c{c}\~ao a um evento, que fun\c{c}\~ao matem\'atica
devemos usar? Podemos obter uma pista sobre tal fun\c{c}\~ao notando que
o que ocorre frequentemente (que possui grande chance de acontecer)
n\~ao nos chama a aten\c{c}\~ao. O nascer do Sol, apesar de bonito e de avisar
que o dia come\c{c}a, n\~ao nos fornece nenhuma informa\c{c}\~ao realmente nova.
Por outro lado, se o c\'eu estiver sem nuvens \`as 14 h e 40 min do dia
14 de dezembro de 2020 em Montevid\'eu, quem perdeu as not\'icias nos
dias anteriores provavelmente vai ficar assustado quando perceber
o dia virando noite, at\'e descobrir que est\'a ocorrendo um eclipse solar.
Esse evento, que n\~ao ocorre frequentemente (\'e pouco prov\'avel), nos
fornece bastante informa\c{c}\~ao pois nos causa surpresa. Este racioc\'inio
nos induz a quantificar a informa\c{c}\~ao, $I$, obtida ao tomarmos conhecimento
da ocorr\^encia de um evento (ou, de forma equivalente, nossa incerteza
sobre o evento antes dele acontecer) usando o rec\'iproco da sua probabilidade
de ocorr\^encia $p$, i.e., $I\equiv 1/p.$

Uma exig\^encia m\'inima que deve ser feita para uma fun\c{c}\~ao que quantifica
informa\c{c}\~ao \'e que ela seja aditiva para eventos independentes. Ou seja,
se $I_{1}$ e $I_{2}$ s\~ao as informa\c{c}\~oes obtidas ao tomarmos conhecimento
dos eventos independentes 1 e 2, se espera que a informa\c{c}\~ao total
obtida seja a soma das duas: $I_{12}=I_{1}+I_{2}$. Tratando esses
eventos como vari\'aveis aleat\'orias, sabemos, da teoria de probabilidades
\cite{prob,DeGroot}, que a probabilidade conjunta de eventos independentes
\'e o produto das probabilidades individuais, ou seja, $p_{12}=p_{1}p_{2}$.
Mas isso implica que a informa\c{c}\~ao total ser\'a: $I_{12}=1/p_{12}=1/p_{1}p_{2}\ne1/p_{1}+1/p_{2}=I_{1}+I_{2}.$
Para obter aditividade da medida de informa\c{c}\~ao, podemos aplicar a
fun\c{c}\~ao logaritmo ao rec\'iproco da probabilidade: 
\begin{equation}
I\equiv\log_{2}(1/p)=\log_{2}p^{-1}=-\log_{2}p.
\end{equation}
\'E f\'acil verificar que essa medida de informa\c{c}\~ao \'e aditiva para eventos
independentes (pois $\log_{2}p_{1}p_{2}=\log_{2}p_{1}+\log_{2}p_{2}$).
Na \'ultima equa\c{c}\~ao e doravante utilizamos o logaritmo na base dois
para medir informa\c{c}\~ao em bits (ao tomarmos conhecimento de um evento
que possui $50\mbox{ \%}$ de chances de ocorrer, i.e., $p=1/2$,
adquirimos $1\mbox{ bit}$ de informa\c{c}\~ao).

Vamos considerar uma vari\'avel aleat\'oria $X$ que pode assumir um conjunto
discreto e finito de valores $\{x_{i}\}_{i=1}^{n}$ com probabilidades
respectivas $\{p_{i}\}_{i=1}^{n}$. Cada um desses valores \'e associado
a um evento. Por exemplo, $X$ pode representar uma moeda, e os dois
eventos seriam a ocorr\^encia de cara ou de coroa em um jogo dessa moeda.
Podemos medir nossa incerteza m\'edia sobre o valor de $X$ usando a
m\'edia da informa\c{c}\~ao dos eventos individuais calculada usando a distribui\c{c}\~ao
de probabilidades $\{p_{i}\}_{i=1}^{n}$: 
\begin{equation}
H(\{p_{i}\})\equiv\sum_{i=1}^{n}p_{i}(-\log_{2}p_{i})=-\sum_{i=1}^{n}p_{i}\log_{2}p_{i}.
\end{equation}
Esta fun\c{c}\~ao, que foi introduzida por Claude Shannon na Ref. \cite{Shannon1948},
\'e denominada \emph{entropia de Shannon} e estabeleceu a base para
a teoria cl\'assica da informa\c{c}\~ao \cite{Cover-Thomas}.

Como os operadores densidade s\~ao operadores hermitianos e positivos
semidefinidos (seus autovalores s\~ao n\'umeros reais maiores ou iguais
a zero) e possuem tra\c{c}o igual a um (o tra\c{c}o de uma matriz \'e a soma
dos elementos na sua diagonal principal), podemos escrever
$\hat{\rho}=\sum_{j=1}^{m}p_{j}|\psi_{j}\rangle\langle\psi_{j}|=\sum_{i=1}^{n}\lambda_{i}|\lambda_{i}\rangle\langle\lambda_{i}|,$
com $\lambda_{i}\in\mathbb{R}$, $\lambda_{i}\ge0$, $\sum_{i=1}^{n}\lambda_{i}=1$
e $\langle\lambda_{i}|\lambda_{j}\rangle=\delta_{ij}$. Ou seja, os
autovalores de $\hat{\rho}$ formam uma distribui\c{c}\~ao de probabilidades.
Ent\~ao, operadores densidade podem ser vistos como uma generaliza\c{c}\~ao
para distribui\c{c}\~oes de probabilidades. Como a fun\c{c}\~ao tra\c{c}o \'e independente
da base usada na representa\c{c}\~ao do operador densidade, a \emph{entropia de von
Neumann}, que foi definida na Ref. \cite{von Neumann}, pode ser escrita
como:
\begin{equation}
S(\hat{\rho})\equiv-\mathrm{Tr}(\hat{\rho}\log_{2}\hat{\rho})=-\sum_{i=1}^{n}\lambda_{i}\log_{2}\lambda_{i},\label{eq:SvN}
\end{equation}
em que $\mathrm{Tr}$ \'e a fun\c{c}\~ao tra\c{c}o. Ou seja, a entropia de von Neumann $S(\hat{\rho})$
\'e a entropia de Shannon da distribui\c{c}\~ao de probabilidades obtida usando
os autovalores de $\hat{\rho}$, i.e., 
\begin{equation}
S(\hat{\rho})=H(\{\lambda_{i}\}).
\end{equation}
Esta \'e uma das fun\c{c}\~oes mais utilizadas atualmente para medir incerteza
sobre o estado de um sistema qu\^antico. Por\'em \'e v\'alido observar que,
embora na teoria qu\^antica da informa\c{c}\~ao \cite{Nielsen_QCQI,Wilde}
exista uma interpreta\c{c}\~ao operacional para essa fun\c{c}\~ao \cite{Schumacher},
crit\'erios mais fortes para quantificar incerteza sobre o operador
densidade s\~ao conhecidos, como por exemplo aquele obtido via teoria
de majora\c{c}\~ao \cite{Nielsen_Maj}. Deixaremos para discutir essas quest\~oes
em uma outra oportunidade.

A entropia de von Neumann \'e comumente escrita na seguinte forma: $S(\hat{\rho})=-k_{B}\mathrm{Tr}(\hat{\rho}\ln\hat{\rho})$, com $k_{B}$ sendo a constante de Boltzmann e $\ln$ o logaritmo natural. Vale observar que esta express\~ao, que aparece naturalmente no contexto da mec\^anica estat\'istica, difere daquela que usamos neste artigo (Eq. (\ref{eq:SvN})) somente por uma constante multiplicativa \cite{Plenio}. Ou seja, n\~ao h\'a nenhuma diferen\c{c}a fundamental entre elas, \'e s\'o a unidade da entropia fornecida que muda. Por isso usamos a  express\~ao (\ref{eq:SvN}), que se encaixa melhor dentro do contexto de teoria de informa\c{c}\~ao qu\^antica, onde recebe um significado operacional em termos n\'umero m\'inimo, m\'edio, de bits qu\^anticos (sistemas qu\^anticos com dois n\'iveis de energia) necess\'arios para representar o estado de um sistema \cite{Schumacher}.

\section{O jogo}

Por simplicidade, vamos considerar um sistema com dois n\'iveis de energia,
cujo espa\c{c}o de estados \'e $\mathbb{C}^{2}$. Das v\'arias escolhas dispon\'iveis,
usaremos part\'iculas de spin 1/2 cujo momento magn\'etico \'e medido utilizando
um aparato de Stern e Gerlach \cite{Sakurai}. Os observ\'aveis que
consideraremos s\~ao as componentes do momento magn\'etico dessas part\'iculas
na dire\c{c}\~ao $z$, $\hat{S}_{z}=(+\hbar/2)|z_{+}\rangle\langle z_{+}|+(-\hbar/2)|z_{-}\rangle\langle z_{-}|$,
e na dire\c{c}\~ao $x$, $\hat{S}_{x}=(+\hbar/2)|x_{+}\rangle\langle x_{+}|+(-\hbar/2)|x_{-}\rangle\langle x_{-}|$,
em que $\hbar$ \'e a constante de Planck,
\begin{equation}
|z_{+}\rangle=\begin{bmatrix}1\\
0
\end{bmatrix}\mbox{, }|z_{-}\rangle=\begin{bmatrix}0\\
1
\end{bmatrix},
\end{equation}
e 
\begin{equation}
|x_{\pm}\rangle=(1/\sqrt{2})(|z_{+}\rangle\pm|z_{-}\rangle).
\end{equation}

As \emph{regras para o jogo} s\~ao as seguintes: Considere um ensemble
de part\'iculas de spin 1/2 preparado no estado $\hat{\rho}=\sum_{j}p_{j}|\psi_{j}\rangle\langle\psi_{j}|$.
Voc\^e poder\'a escolher um \'unico observ\'avel do sistema para ser medido
e precisar\'a escolher tamb\'em um \'unico autovetor desse observ\'avel. Cada
vez que o valor do observ\'avel correspondente a este autovetor for
obtido voc\^e ganha um ponto, caso contr\'ario voc\^e perde um ponto. O
objetivo \'e obter o maior saldo de pontos, $N_{p}$, poss\'ivel para
um n\'umero muito grande, $N$, de medidas realizadas. Queremos relacionar
sua pontua\c{c}\~ao com a entropia de von Neumann, $S(\hat{\rho})$. Para
calcul\'a-la, voc\^e precisar\'a escolher uma base, por exemplo $\{|z_{+}\rangle,|z_{-}\rangle\}$,
obter a representa\c{c}\~ao matricial de $\hat{\rho}$ nessa base ($\rho_{11}=\langle z_{+}|\hat{\rho}|z_{+}\rangle$,
$\rho_{12}=\langle z_{+}|\hat{\rho}|z_{-}\rangle$, $\rho_{21}=\langle z_{-}|\hat{\rho}|z_{+}\rangle$
e $\rho_{22}=\langle z_{-}|\hat{\rho}|z_{-}\rangle$) e diagonalizar
essa matriz $2\mathrm{x}2$. Feito isso, precisar\'as ainda somar o produto
dos autovalores de $\hat{\rho}$ pelo seu logaritmo, como na Eq. (\ref{eq:SvN}).
Se $\hat{\rho}$ possuir um autovalor nulo, usamos $\lim_{x\rightarrow0}x\log_{2}x=0$
para justificar $0\log_{2}0=0$. Com o objetivo de criar intui\c{c}\~ao
sobre o problema, vamos considerar alguns casos particulares de ensembles
preparados:

$\bullet$ $\hat{\rho}_{1}=|z_{+}\rangle\langle z_{+}|$\\Nesse caso,
como voc\^e sabe que todas as part\'{i}culas do ensemble est\~ao no mesmo
estado, tudo fica mais f\'acil. Voc\^e pode escolher medir o observ\'avel
$\hat{S}_{z}$ e pode escolher como seu autovetor o autoestado $|z_{+}\rangle$
de $\hat{S}_{z}$. Como a probabilidade de obter o valor $+\hbar/2$
para $\hat{S}_{z}$ \'e um para um sistema preparado no estado $|z_{+}\rangle$,
voc\^e ganha em todas as medidas e faz um saldo de $N_{p}=N-0=N$ pontos,
o m\'aximo poss\'ivel. Isso \'e esperado pois n\~ao h\'a incerteza alguma sobre
a prepara\c{c}\~ao do sistema, o que \'e confirmado pelo valor da entropia
de von Neumann desse estado, $S(\hat{\rho}_{1})=S(|z_{+}\rangle)=-1\log_{2}1-0\log_{2}0=0.$

Note que, como ocorreu para este ensemble, a entropia de von Neumann \'e nula para qualquer estado puro, i.e., $S(|\psi\rangle)=0$. Entretanto, isso n\~ao significa que podemos prever o resultado de qualquer medida que possa ser feita nesse sistema, mas sim que existe um observ\'avel cujo valor \'e totalmente previs\'ivel, pois est\'a definido antes que testes sejam feitos (seu valor \'e o seu autovalor correspondente ao autovetor $|\psi\rangle$).

\begin{figure}
\includegraphics[scale=1]{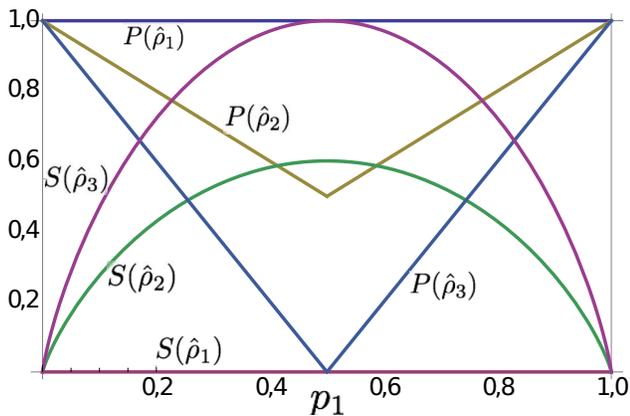}
\protect\caption{Entropia de von Neumann, $S$, e pontua\c{c}\~ao m\'axima normalizada, $P\equiv N_{p}/N$,
para os ensembles $\hat{\rho}_{1}$, $\hat{\rho}_{2}$ e $\hat{\rho}_{3}$
como fun\c{c}\~ao da probabilidade de prepara\c{c}\~ao $p_{1}=1-p_{2}$.}
\label{fig1}
\end{figure}

$\bullet$ $\hat{\rho}_{2}=p_{1}|z_{+}\rangle\langle z_{+}|+p_{2}|x_{+}\rangle\langle x_{+}|$\\Neste
exemplo, \'e aconselh\'avel analisar com cuidado o estado preparado antes
de fazer suas escolhas. Veja que existe incerteza cl\'assica sobre se
$|z_{+}\rangle$ ou se $|x_{+}\rangle$ foi preparado. Al\'em disso,
esses estados s\~ao autovetores de observ\'aveis incompat\'iveis, o que
implica na exist\^encia de incerteza qu\^antica (pois voc\^e pode escolher
um \'unico observ\'avel para medir). Para facilitar sua escolha, use o
fato de que $|x_{+}\rangle$ \'e uma superposi\c{c}\~ao dos autovetores de
$\hat{S}_{z}$. Assim $\langle z_{\pm}|x_{+}\rangle=(1/\sqrt{2})(\langle z_{\pm}|z_{+}\rangle+\langle z_{\pm}|z_{-}\rangle)=1/\sqrt{2}.$
Usando a regra de Born vemos que, para as part\'iculas preparadas no
estado $|x_{+}\rangle$, a probabilidade de medir $+\hbar/2$ ou $-\hbar/2$
para $\hat{S}_{z}$ \'e 1/2. Por conseguinte, se voc\^e escolher medir
$\hat{S}_{z}$ e escolher $|z_{+}\rangle$, seu n\'umero de pontos ser\'a
$N_{p}=p_{1}N+p_{2}N(1/2)-p_{2}N(1/2)=p_{1}N.$ O primeiro termo dessa
equa\c{c}\~ao aparece pois para todos os membros do ensemble preparados
no estado $|z_{+}\rangle$ voc\^e ganha. O segundo e terceiro termos
s\~ao obtidos notando que para as part\'iculas preparadas no estado $|x_{+}\rangle$
voc\^e ganha metade das vezes e perde em metade das medidas. De forma
an\'aloga, podemos usar $|z_{+}\rangle=(1/\sqrt{2})(|x_{+}\rangle+|x_{-}\rangle)$
para verificar que se $\hat{S}_{x}$ e $|x_{+}\rangle$ forem escolhidos,
ent\~ao $N_{p}=p_{1}N(1/2)-p_{1}N(1/2)+p_{2}N=p_{2}N.$ A diferen\c{c}a
entre os $N_{p}$ das duas escolhas \'e $N(p_{1}-p_{2})$. Assim, usamos
o vi\'es nas probabilidades, ou seja, verificamos se $p_{1}-p_{2}$
\'e maior ou menor que zero, para fazer nossa escolha de forma apropriada
a maximizar $N_{p}$. Por exemplo, se $p_{1}=3/4$ e $p_{2}=1/4$
a primeira escolha nos fornece $N_{p}=0,75N$ e a segunda escolha
resulta em $N_{p}=0,25N$. Por outro lado, se $p_{1}=p_{2}=1/2$ as
duas escolhas resultar\~ao em $N_{p}=0,5N$. Esse ser\'a nosso pior saldo
de pontos, pois a incerteza cl\'assica \'e m\'axima nesse caso. Para este
ensemble a entropia de von Neumann \'e dada por $S(\hat{\rho}_{2})=-\lambda_{1}\log_{2}\lambda_{1}-(1-\lambda_{1})\log_{2}(1-\lambda_{1}),$
com $\lambda_{1}=(1-\sqrt{1-2p_{1}p_{2}})/2$, e est\'a mostrada na
Fig. \ref{fig1}.

$\bullet$ $\hat{\rho}_{3}=p_{1}|z_{+}\rangle\langle z_{+}|+p_{2}|z_{-}\rangle\langle z_{-}|$\\Este
ensemble \'e obtido trocando-se $|x_{+}\rangle$ por $|z_{-}\rangle$
no \'ultimo exemplo. Portanto a an\'alise \'e, de certa forma, parecida.
Nesse caso \'e natural escolher $\hat{S}_{z}$ para ser medido. Ent\~ao,
se o autovetor escolhido for $|z_{+}\rangle$, o saldo de pontos ser\'a
$N_{p}=p_{1}N-p_{2}N=(2p_{1}-1)N.$ Se, no entanto, escolhermos $|z_{-}\rangle$,
ent\~ao $N_{p}=p_{2}N-p_{1}N=(2p_{2}-1)N.$ A diferen\c{c}a entre essas
express\~oes para $N_{p}$ \'e $2(p_{1}-p_{2})N$. Usamos o sinal desse
n\'umero para escolher o autovetor que maximiza nossa pontua\c{c}\~ao. Aqui
\'e interessante considerar o caso de equiprobabilidade $p_{1}=p_{2}=1/2$,
para o qual $N_{p}=0$. Note que nossa pontua\c{c}\~ao nesse caso, al\'em
de ser menor quando comparada com o mesmo caso no exemplo anterior,
\'e a m\'inima poss\'ivel. A entropia de von Neumann desse ensemble \'e: $S(\hat{\rho}_{3})=-p_{1}\log_{2}p_{1}-(1-p_{1})\log_{2}(1-p_{1}),$
e possui o m\'aximo valor poss\'ivel quando $p_{1}=1/2$ (veja a Fig.
\ref{fig1}).

A entropia de von Neumann, $S$, e o m\'aximo saldo de pontos normalizado,
$P\equiv N_{p}/N$, est\~ao mostrados na Fig. \ref{fig1} para os tr\^es
ensembles considerados. Note que, para um dado ensemble, $P$ diminui (aumenta)
com o aumento (decr\'ecimo) de $S$. Al\'em disso, o aumento de $S$ de $\hat{\rho}_{3}$
(que \'e uma mistura estat\'istica de estados ortogonais) em rela\c{c}\~ao a
$\hat{\rho}_{2}$ (que \'e uma mistura estat\'istica de estados n\~ao ortogonais)
est\'a associado a uma diminui\c{c}\~ao relativa de $P$. Vale observar que
$S$ depende s\'o do estado do sistema, enquanto que $P$ depende tamb\'em
do observ\'avel medido e do autovalor escolhido. Mas a principal mensagem
que gostar\'iamos de passar com esses exemplos \'e que o seu saldo de
pontos est\'a relacionado com a sua capacidade de, usando a informa\c{c}\~ao
que tens sobre como o sistema foi preparado, prever o que acontecer\'a
em uma medida de um observ\'avel desse sistema. Por sua vez, a falta
dessa informa\c{c}\~ao, ou seja, a incerteza sobre o estado do sistema,
que pode ser quantificada usando a entropia de von Neumann, diminuir\'a
sua capacidade de previs\~ao e por consequ\^encia o seu saldo de pontos. 

Com esse jogo fica claro que o que \'e incerteza qu\^antica quando o observ\'avel
medido \'e incompat\'ivel com o observ\'avel usado na prepara\c{c}\~ao do sistema,
na presen\c{c}a de incerteza cl\'assica torna-se informa\c{c}\~ao que pode ser
utilizada para aumentar nossa pontua\c{c}\~ao no jogo, ou seja, para diminuir
nossa incerteza sobre o estado do sistema e aumentar nossa capacidade
de previs\~ao concernindo resultados de testes. Essas observa\c{c}\~oes podem
ser generalizadas para sistemas discretos com um n\'umero finito $d$
de n\'iveis de energia. Nesse caso, se $\{|o_{i}\rangle\}$ s\~ao autovetores
do observ\'avel $\hat{O}$, $S(\sum_{i=1}^{d}(1/d)|o_{i}\rangle\langle o_{i}|)=\log_{2}d$
\'e o valor m\'aximo que a entropia de von Neumann pode assumir. A presen\c{c}a
de um ou mais estados que n\~ao s\~ao autovetores de $\hat{O}$ nessa mistura diminuir\'a
o valor de $S$.

\section{Considera\c{c}\~oes finais}

Nesse artigo apresentamos os postulados da mec\^anica qu\^antica de forma
orientada a facilitar o entendimento da entropia de von Neumann. Um
jogo simples foi proposto para ajudar nessa tarefa. Com esse jogo
enfatizamos o papel distinto da incerteza cl\'assica, que est\'a relacionada
com a falta de informa\c{c}\~ao sobre a prepara\c{c}\~ao do sistema, e da incerteza
qu\^antica, que existe quando medimos um observ\'avel para um sistema
preparado em um estado que n\~ao \'e seu autovetor. Esperamos que essa
exposi\c{c}\~ao ajude tamb\'em na compreens\~ao de v\'arios outros t\'opicos interessantes
relacionados \`a entropia de von Neumann, como por exemplo a sua interpreta\c{c}\~ao
operacional \cite{Schumacher} e seus v\'inculos em sistemas qu\^anticos
compostos \cite{ESSA}.

\renewcommand\acknowledgmentsname{Agradecimentos}

\begin{acknowledgments}
Esse trabalho foi parcialmente financiado pelo Conselho Nacional de
Desenvolvimento Cient\'ifico e Tecnol\'ogico (CNPq), via o Instituto Nacional
de Ci\^encia e Tecnologia de Informa\c{c}\~ao Qu\^antica (INCT-IQ), e pela Coordena\c{c}\~ao de Aperfei\c{c}oamento de Pessoal de N\'ivel Superior ($\mathrm{CAPES}$). Agradecemos ao revisor pelas suas observa\c{c}\~oes pertinentes e cooperantes.  
\end{acknowledgments}

\end{document}